\RequirePackage[l2tabu, orthodox]{nag}
\documentclass[12pt]{article}

\usepackage[nonatbib,final]{nips_2016}

\usepackage{setspace}
\usepackage{makecell}
\newcolumntype{M}[1]{>{\centering\arraybackslash}m{#1}}
\usepackage[T1]{fontenc}

\usepackage{tgtermes}
\usepackage{amsmath}
\usepackage{scalefnt,letltxmacro}
\LetLtxMacro{\oldtextsc}{\textsc}
\renewcommand{\textsc}[1]{\oldtextsc{\scalefont{1.10}#1}}
\usepackage[scaled=0.92]{PTSans}
\usepackage{inconsolata}
\usepackage{mathbbol}
\usepackage{amsthm}
\usepackage{amssymb}

\usepackage{setspace}

\usepackage[usenames,dvipsnames]{xcolor}
\definecolor{shadecolor}{gray}{0.9}

\usepackage[english]{babel}
\usepackage{afterpage}
\usepackage{framed}
\usepackage{nicefrac}

{\endMakeFramed}

\DeclareRobustCommand{\parhead}[1]{\textbf{#1}~}

\usepackage{lineno}

\usepackage{ragged2e}

\usepackage{marginnote}

\newcounter{parcount}

\usepackage{graphicx}
\usepackage[labelfont=bf]{caption}
\usepackage[format=hang]{subcaption}
\usepackage{wrapfig}

\usepackage{booktabs}
\usepackage{multirow}
\usepackage{longtable}

\usepackage{natbib}
\usepackage{bibunits}

\usepackage[algoruled]{algorithm2e}
\usepackage{listings}
\usepackage{fancyvrb}
\fvset{fontsize=\normalsize}

\SetKwComment{Comment}{$\triangleright$\ }{}

\usepackage[colorlinks,linktoc=all]{hyperref}
\usepackage[all]{hypcap}
\hypersetup{citecolor=Violet}
\hypersetup{linkcolor=black}
\hypersetup{urlcolor=MidnightBlue}

\usepackage[nameinlink]{cleveref}

\usepackage[acronym,smallcaps,nowarn]{glossaries}

\usepackage{listings}
\usepackage{lstbayes}
\lstdefinestyle{mystyle}{
    commentstyle=\color{OliveGreen},
    numberstyle=\tiny\color{black!60},
    stringstyle=\color{BrickRed},
    basicstyle=\ttfamily\scriptsize,
    breakatwhitespace=false,
    breaklines=true,
    captionpos=b,
    keepspaces=true,
    numbers=none,
    numbersep=5pt,
    showspaces=false,
    showstringspaces=false,
    showtabs=false,
    tabsize=2
}
\usepackage{enumitem}

\usepackage{centernot}

\DeclareRobustCommand{\mb}[1]{\ensuremath{\mathbf{\boldsymbol{#1}}}}

\DeclareMathOperator*{\argmax}{arg\,max}

\crefname{lemma}{lemma}{lemmas}
\Crefname{lemma}{Lemma}{Lemmas}
\crefname{thm}{theorem}{theorems}
\Crefname{thm}{Theorem}{Theorems}
\crefname{prop}{proposition}{propositions}
\Crefname{prop}{Proposition}{Propositions}

\creflabelformat{equation}{#1#2#3}
\crefname{equation}{eq.}{eqs.}
\Crefname{equation}{Eq.}{Eqs.}
\Crefname{section}{\S}{\S}

\newtheorem{thm}{Theorem} 

\newcommand\independent{\protect\mathpalette{\protect\independenT}{\perp}}
\def\independenT#1#2{\mathrel{\rlap{$#1#2$}\mkern2mu{#1#2}}}

\newcommand{\mba}{\mb{a}}

\newcommand{\cN}{\mathcal{N}}

\newcommand{\g}{\, | \,}

\newcommand{\E}[2]{\mathbb{E}_{#1}\left[#2\right]}

\usepackage{booktabs,arydshln}
\makeatletter
\def\adl@drawiv#1#2#3{%
        \hskip.5\tabcolsep
        \xleaders#3{#2.5\@tempdimb #1{1}#2.5\@tempdimb}%
                #2\z@ plus1fil minus1fil\relax
        \hskip.5\tabcolsep}
\newcommand{\cdashlinelr}[1]{%
  \noalign{\vskip\aboverulesep
           \global\let\@dashdrawstore\adl@draw
           \global\let\adl@draw\adl@drawiv}
  \cdashline{#1}
  \noalign{\global\let\adl@draw\@dashdrawstore
           \vskip\belowrulesep}}
\makeatother

\newacronym{ELBO}{elbo}{evidence lower bound}
\newacronym{GMM}{gmm}{Gaussian mixture model}
\newacronym{KL}{kl}{Kullback-Leibler}
\newacronym{LDA}{lda}{latent Dirichlet allocation}
\newacronym{SVI}{svi}{stochastic variational inference}

\newacronym{MLE}{mle}{maximum likelihood estimate}
\newacronym{MCMC}{mcmc}{Markov chain Monte Carlo}
\newacronym{HMC}{hmc}{Hamiltonian Monte Carlo}

\newacronym{LBFGS}{l-bfgs}{limited-memory Broyden-Fletcher-Goldfarb-Shanno}
\newacronym{ADVI}{advi}{automatic differentiation variational inference}
\newacronym{NUTS}{nuts}{No-U-Turn sampler}

\newacronym{GLM}{glm}{generalized linear model}

\newacronym{IF}{if}{influence function}

\newacronym{PF}{pf}{Poisson factorization}

\newacronym[\glsshortpluralkey={rpm}]
{RPM}{rpm}{reweighted probabilistic model}

\newacronym{NDCG}{ndcg}{normalized discounted cumulative gain}

\newacronym{MAP}{map}{mean average precision}
\newacronym{MSE}{mse}{mean squared error}
\newacronym{IPW}{ipw}{inverse propensity weighting}

\usepackage{tikz}
\usetikzlibrary{bayesnet}
\usepackage{pgfplots}
\pgfplotsset{compat=newest}
\pgfplotsset{plot coordinates/math parser=false}
\usepgfplotslibrary{statistics}

\pgfdeclarelayer{edgelayer}
\pgfdeclarelayer{nodelayer}
\pgfsetlayers{edgelayer,nodelayer,main}

\definecolor{hexcolor0xbfbfbf}{rgb}{0.749,0.749,0.749}

\tikzset{>=latex}
\tikzstyle{none}   = [inner sep=0pt]
\tikzstyle{line}   = [ thick, -, shorten <=1pt, shorten >=1pt ]
\tikzstyle{arrow}  = [ thick,  ->, shorten <=1pt, shorten >=1pt ]
\tikzstyle{ardash} = [ thick dotted, ->, shorten <=1pt, shorten >=1pt ]

\tikzstyle{empty}=[circle,opacity=0.0,text opacity=1.0,minimum width=4pt,minimum height=4pt]
\tikzstyle{box}=[rectangle,fill=White,draw=Black]
\tikzstyle{filled}=[circle,fill=hexcolor0xbfbfbf,draw=Black]
\tikzstyle{hollow}=[circle,fill=White,draw=Black]
\tikzstyle{param}=[rectangle,fill=Black,draw=Black,inner sep=0pt,minimum width=4pt,minimum height=4pt]
\tikzstyle{paramhollow}=[rectangle,fill=White,draw=Black,inner sep=0pt,minimum
width=4pt,minimum height=4pt]

\usepackage{times}
\usepackage{adjustbox} 
\usepackage{tcolorbox}
\usepackage{cuted}

\everypar=\expandafter{\the\everypar\loosness=-1 }
\title{The Deconfounded Recommender: \\
A Causal Inference Approach to Recommendation}

\author{
  Yixin Wang\\
  Columbia University\\
  \And
  Dawen Liang\\
  Netflix Inc.\\
  \And
  Laurent Charlin\thanks{Canada CIFAR AI Chair}\\
  Mila, HEC Montr\'eal\\
  \And
  David M.~Blei\\
  Columbia University\\
}

\begin{document}

\maketitle

\begin{bibunit}[alp]

\begin{abstract}
  The goal of recommendation is to show users items that they will
  like. Though usually framed as a prediction, the spirit of
  recommendation is to answer an interventional question---for each
  user and movie, what would the rating be if we ``forced'' the user
  to watch the movie?  To this end, we develop a causal approach to
  recommendation, one where watching a movie is a ``treatment'' and a
  user's rating is an ``outcome.''  The problem is there may be
  unobserved confounders, variables that affect both which movies the
  users watch and how they rate them; unobserved confounders impede
  causal predictions with observational data. To solve this problem,
  we develop the deconfounded recommender, a way to use classical
  recommendation models for causal recommendation.  Following
  \citet{wang2018blessings}, the deconfounded recommender involves two
  probabilistic models. The first models which movies the users watch;
  it provides a substitute for the unobserved confounders. The second
  one models how each user rates each movie; it employs the substitute
  to help account for confounders.  This two-stage approach removes
  bias due to confounding. It improves recommendation and enjoys
  stable performance against interventions on test sets.
\end{abstract}


\section{Introduction}
\label{sec:introduction}

The goal of a recommender is to show its users items that they will
like.  Given a dataset of users' ratings, a recommender system learns
the preferences of the users, predicts the users' ratings on those
items they did not rate, and finally makes suggestions based on those
predictions. In this paper we develop an approach to recommendation
based on causal inference.

Why is recommendation a causal inference?  Concretely, suppose the
items are movies and the users rate movies they have seen. In
prediction, the recommender system is trying to answer the question
``How would the user rate this movie if he or she saw it?''  But this
is a question about an \textit{intervention}: what would the rating be
if we ``forced'' the user to watch the movie? One tenet of causal
inference is that predictions under intervention are different from
usual ``out-of-sample'' predictions.

Framing recommendation as a causal problem differs from the
traditional approach. The traditional approach builds a model from
observed ratings data, often a matrix factorization, and then uses
that model to predict unseen ratings. But this strategy only provides
valid causal inferences---in the intervention sense described
above---if users randomly watched and rated movies. (This is akin to a
randomized clinical trial, where the treatment is a movie and the
response is a rating.)

But users do not (usually) watch movies at random and, consequently,
answering the causal question from observed ratings data is
challenging.  The issue is that there may be \textit{confounders},
variables that affect both the treatment assignments (which movies the
users watch) and the outcomes (how they rate them).  For example,
users pick movies by directors they like and then tend to like those
movies. The director is a confounder that biases our inferences.
Compounding this issue, the confounders might be difficult (or
impossible) to measure.  Further, the theory around causal inferences
says that these inferences are valid only if we have accounted for all
confounders \citep{rosenbaum1983central}. And, alas, whether we have
indeed measured all confounders~is~uncheckable~\citep{holland1985statistics}.

How can we overcome these obstacles?  In this paper, we develop the
\textit{deconfounded recommender}, a method that tries to correct
classical matrix factorization for unobserved confounding.  The
deconfounded recommender builds on the two sources of information in
recommendation data: which movies each user decided to watch and the
user's rating for each of those movies. It posits that the two types
of information come from different models---the \textit{exposure} data
comes from a model by which users discover movies to watch; the
\textit{ratings} data comes from a model by which users decide which
movies they like. The ratings data entangles both types of
information---users only rate movies that they see---and so classical
matrix factorization is biased by the exposure model, i.e., that users
do not randomly choose movies.

The deconfounded recommender tries to correct this bias. It first uses
the exposure data to estimate a model of which movies each user is
likely to consider.  (In the language of recommender systems, the
exposure data is a form of ``implicit'' data.)  It then uses this
exposure model to estimate a substitute for the unobserved
confounders.  Finally, it fits a ratings model (e.g., matrix
factorization) that accounts for the substitute confounders.  The
justification for this approach comes from \citet{wang2018blessings};
correlations among the considered movies provide indirect evidence for
confounders.\footnote{The deconfounded recommender focuses on how the
  exposure of each individual movie (i.e. one of the many causes)
  causally affects its observed rating (\Cref{eq:dcfrec}); we rely on
  Theorem 7 of \citet{wang2018blessings} for the identification of
  causal parameters. Note this result does not contradict the causal
  non-identification examples given in \citet{d2019multi}, which
  operate under different assumptions.}

Consider a film enthusiast who mostly watches western action movies
but who has also enjoyed two Korean dramas, even though non-English
movies are not easily accessible in her area.  A traditional
recommender will infer preferences that center around westerns; the
dramas carry comparatively little weight.  The deconfounded
recommender will also detect the preference for westerns, but it will
further up-weight the preference for Korean dramas.  The reason is
that the history of the user indicates that she is unlikely to have
been exposed to many non-English movies, and she liked the two Korean
dramas that she did see.  Consequently, when recommending from among
the unwatched movies, the deconfounded recommender promotes other
Korean dramas along with westerns.

Below we develop the deconfounded recommender.  We empirically study
it on both simulated data, where we control the amount of confounding,
and real data, about shopping and movies.  (Software that replicates
the empirical studies is provided in the supplementary material.)
Compared to existing approaches, its performance is more robust to
unobserved confounding; it predicts the ratings better and
consistently improves recommendation.

\parhead{Related work.} This work draws on several threads of previous
research.

The first is on evaluating recommendation algorithms via biased
data. It is mostly explored in the multi-armed bandit literature
\citep{li2015counterfactual, zhao2013interactive,
  vanchinathan2014explore, li2010contextual}. These works focus on
online learning and rely on importance sampling. Here we consider an
orthogonal problem.  We reason about user preferences, rather than
recommendation algorithms, and we use offline learning and parametric
models.

The second thread is around the missing-not-completely-at-random
assumption in recommendation methods.  \citet{marlin2009collaborative}
studied the effect of violating this assumption in ratings.  Similar
to our exposure model, they posit an explicit missingness model that
leads to improvements in predicting ratings. Later, other researchers
proposed different rating models to accommodate this violated
assumption \citep{liang2016modeling, hernandez2014probabilistic,
  ling2012response}.  In contrast to these works, we take an
explicitly causal view of the problem.  While violating the
missing-not-completely-at-random assumption is one form of confounding
bias~\citep{ding2018causal}, the causal view opens up the door to
other debiasing tools, such as the
deconfounder~\citep{wang2018blessings}.

Finally, the recent work of \citet{schnabel2016recommendations} also
adapted causal inference---\gls{IPW}, in
particular---to address missingness. Their propensity models rely on
either observed ratings of a randomized trial or externally observed
user and item covariates. In contrast, our work relies solely on the
observed ratings: we do not require ratings from a gold-standard
randomized exposure nor do we use external covariates. In
\Cref{sec:empirical}, we show that the deconfounded recommender
provides better recommendations than
\citet{schnabel2016recommendations}.


\section{The deconfounded recommender}
\label{sec:vanilladeconfounder}

We frame recommendation as a causal inference and develop the
deconfounded recommender.

\parhead{Matrix factorization as potential outcomes.}  We first set up
notation. Denote $a_{ui}$ as the indicator of whether user $u$ rated
movie $i$.  Let $y_{ui}(1)$ be the rating that user $u$ would give
movie $i$ if she watches it. This rating is observed if the user $u$
watched and rated the movie $i$; otherwise it is unobserved. Similarly
define $y_{ui}(0)$ to be the rating of user $u$ on movie $i$ if she
does not see the movie. (We often ``observe'' $y_{ui}(0) = 0$ in
recommendation data; unrated movie entries are filled with zeros.)
The pair $(y_{ui}(0), y_{ui}(1))$ is the \textit{potential outcomes}
notation in the Rubin causal model \citep{imbens2015causal,
  rubin1974estimating, rubin2005causal}, where watching a movie is a
``treatment'' and a user's rating of the movie is an ``outcome.''

A recommender system observes users' ratings of movies. We can think
of these observations as two datasets. One dataset contains (binary)
exposures, $\{a_{ui}, u = 1, \ldots, U, i = 1, \ldots, I\}$. It
indicates who watched what. The other dataset contains the ratings for
movies that users watched,
$\{y_{ui}(a_{ui})\text{ for }(u,i)\text{ such that }a_{ui}=1 \}$.

The goal of the system is to recommend movies its users will like. It
first estimates $y_{ui}(1)$ for user-movie pairs such that $a_{ui}=0$;
that is, it predicts each user's ratings for their unseen movies. It
then uses these estimates to suggest movies to each user.  Note the
estimate of $y_{ui}(1)$ is a prediction under intervention: ``What
would the rating be if user $u$ was forced to see movie $i$?''

To form the prediction of $y_{ui}(1)$, we recast matrix factorization
in the potential outcomes framework. First set up an \textit{outcome
  model},
\begin{align}
  \label{eq:matrix-factorization}
  y_{ui}(a) = \theta_u^\top\beta_i\cdot a+\epsilon_{ui},\quad
  \epsilon_{ui}\sim\cN(0,\sigma^2).
\end{align}
When $a=1$ (i.e., user $u$ watches movie $i$), this model says that
the rating comes from a Gaussian distribution whose mean combines user
preferences $\theta_u$ and item attributes $\beta_i$. When $a=0$, the
``rating'' is a zero-mean Gaussian.

Fitting \Cref{eq:matrix-factorization} to the observed data recovers
classical probabilistic matrix factorization
\citep{mnih2008probabilistic}. Its log likelihood, as a function of
$\theta$ and $\beta$, only involves observed ratings; it ignores the
unexposed items. The fitted model can then predict
$\E{}{y_{ui}(1)} = \theta_u^\top\beta_i$ for every (unwatched)
user-movie pair. These predictions suggest movies that users would
like.

\parhead{Classical causal inference and adjusting for confounders in
  recommendation.}  But matrix factorization does not provide an
unbiased causal inference of $y_{ui}(1)$.  The theory around potential
outcomes says we can only estimate $y_{ui}(1)$ if we assume
\textit{ignorability}. For all users $u$, ignorability requires
$\{\boldsymbol{y}_{u}(0), \boldsymbol{y}_u(1)\}\independent \mba_u$,
where $\boldsymbol{y}_u(a) = (y_{u1}(a), \ldots, y_{uI}(a))$ and
$\mba_u = (a_{u1}, \ldots, a_{uI})$. In words, the vector of movies a
user watches $\mba_u$ is independent of how she would rate them if she
watched them all $\boldsymbol{y}_u(1)$ (and if she watched none
$\boldsymbol{y}_{u}(0)$).

Ignorability clearly does not hold for $\boldsymbol{y}_u(1)$---the
process by which users find movies is not independent of how they rate
them. Practically, this violation biases the estimates of user
preferences $\theta_u$: movies that $u$ is not likely to see are
downweighted and vice versa. Again consider the American user who
enjoyed two Korean dramas and rated them highly.  Because she has only
two high ratings of Korean dramas in the data, her preference for
Korean dramas carries less weight than her other ratings; it is biased
downward. Biased estimates of preferences lead to biased predictions
of ratings.

When ignorability does not hold, classical causal inference asks us to
measure and control for confounders
\citep{rubin2005causal,pearl2009causality}. These are variables that
affect both the exposure and the ratings. Consider the location of a
user as an example. It affects both which movies they are exposed to
and (perhaps) what kinds of movies they tend to like.

Suppose we measured these per-user confounders $w_u$; they satisfy
$\{\boldsymbol{y}_{u}(0), \boldsymbol{y}_u(1)\}\independent \mba_u \g w_u$. Classical causal
inference controls for them in the outcome model,
\begin{align}
  \label{eq:caupmf}
  y_{ui}(a) = \theta_u^\top\beta_i\cdot a+\eta^\top w_u +\epsilon_{ui},\quad
  \epsilon_{ui}\sim\cN(0,\sigma^2).
\end{align}
However, this solution requires we measure \textit{all} confounders.
This assumption is known as \textit{strong ignorability}.\footnote{In
causal graphical models, this requirement is equivalent to ``no open
backdoor paths'' \citep{pearl2009causality}.} Unfortunately, it is
untestable \citep{holland1985statistics}.

\parhead{The deconfounded recommender.}  We now develop the
deconfounded recommender. It leverages the dependencies among the
exposure (``which movies the users watch'') as indirect evidence for
unobserved confounders. It uses a model of the exposure to construct a
substitute confounder; it then conditions on the substitute when
modeling the ratings.

The key idea is that causal inference for recommendation systems is a
\textit{multiple causal inference}: there are multiple treatments.
Each user's binary exposure to each movie $a_{ui}$ is a treatment;
thus there are $I$ treatments for each user. The vector of ratings
$\boldsymbol{y}_u(1)$ is the outcome; this is an $I$-vector, which is
partially observed.  The multiplicity of treatments enables causal
inference with unobserved confounders \citep{wang2018blessings}.

\glsreset{PF}

The first step is to fit a model to the exposure data. We use \gls{PF}
model \citep{gopalan2015scalable}. \gls{PF} assumes the data come from
the following process,
\begin{align}
&\pi_u\stackrel{iid}{\sim}\textrm{Gamma}(c_1, c_2),\qquad
\lambda_i\stackrel{iid}{\sim}\textrm{Gamma}(c_3, c_4), \\
&a_{ui}\,|\,\pi_u, \lambda_i \sim \textrm{Poisson}(\pi_u^\top
\lambda_i), \qquad\qquad\forall u, i,
\label{eq:pf}
\end{align}
where both $\pi_u$ and $\lambda_i$ are nonnegative $K$-vectors. The user
factor $\pi_u$ captures user preferences (in picking what movies to
watch) and the item vector $\lambda_i$ captures item attributes. \gls{PF} is
a scalable variant of nonnegative factorization and is especially
suited to binary data \citep{gopalan2015scalable}.  It is fit with
coordinate ascent variational inference, which scales with the number
of nonzeros in $\mba = \{a_{ui}\}_{u,i}$.\footnote{While the Bernoulli
  distribution is more natural to model binary exposure, \gls{PF} is
  more computationally efficient than Bernoulli factorization and
  there are several precedents to modeling binary data with a Poisson
  \citep{gopalan2015scalable,gopalan2014content}.  \gls{PF} scales
  linearly with the number of \emph{nonzero} entries in the exposure
  matrix $\{a_{ui}\}_{U\times I}$ while Bernoulli scales with the
  number of \emph{all} entries.  Further, the Poisson distribution
  closely approximates the Bernoulli when the exposure matrix
  $\{a_{ui}\}_{U\times I}$ is sparse
  \citep{degroot2012probability}. Finally, \gls{PF} can also model
  non-binary \emph{count} exposures: e.g., \gls{PF} can model
  exposures that count how many times a user has been exposed to an
  item.}

With a fitted \gls{PF} model, the deconfounded recommender computes a
substitute for unobserved confounders. It reconstructs the exposure
matrix $\hat{a}$ from the \gls{PF} fit,
\begin{align}
\hat{a}_{ui} = \mathbb{E}_{\gls{PF}}[\pi_u^\top \lambda_i~|~\mba],
\label{eq:reconstr}
\end{align}
where $\mba$ is the observed exposure for all users, and the
expectation is taken over the posteriors computed from the \gls{PF}
model. This is the posterior predictive mean of $\pi_u^\top \lambda_i$. The
reconstructed exposure $\hat{a}_{ij}$ serves as a substitute
confounder \citep{wang2018blessings}.

Finally, the deconfounded recommender posits an outcome model
conditional on the substitute confounders $\hat{a}$,
\begin{align}
  y_{ui}(a) = \theta_u^\top\beta_i\cdot a+\gamma_u\cdot
  \hat{a}_{ui}+\epsilon_{ui}, \quad\epsilon_{ui}\sim\cN(0,\sigma^2),
  \label{eq:dcfout}
\end{align}
where $\gamma_u$ is a user-specific coefficient that describes how
much the substitute confounder $\hat{a}$ contributes to the
ratings. The deconfounded recommender fits this outcome model to the
observed data; it infers $\theta_u, \beta_i, \gamma_u$, via
\textit{maximum a posteriori} estimation.  Note in fitting
\Cref{eq:dcfout}, the coefficients $\theta_u, \beta_i$ are fit only
with the observed user ratings (i.e., $a_{ui}$ = 1) because $a_{ui}=0$
zeroes out the term that involves them; in contrast, the coefficient
$\gamma_u$ is fit to all movies (both $a_{ui}$ = 0 and $a_{ui}$ = 1)
because $\hat{a}_{ui}$ is always non-zero.

To form recommendations, the deconfounded recommender calculates all
the potential ratings $y_{ui}(1)$ with the fitted $\hat{\theta}_u,
\hat{\beta}_i, \hat{\gamma}_u$. It then orders the potential ratings
of the unseen movies.  These are causal recommendations.
\Cref{alg:deconfounder} provides the algorithm for forming
recommendations with the deconfounded recommender.

\parhead{Why does it work?}  \glsreset{PF} \gls{PF} learns a per-user
latent variable $\pi_u$ from the exposure matrix $a_{ui}$, and we take
$\pi_u$ as a substitute confounder.  What justifies this approach is
that \gls{PF} admits a special conditional independence structure:
conditional on $\pi_u$, the treatments $a_{ui}$ are independent
(\Cref{eq:pf}). If the exposure model \gls{PF} fits the data well,
then the per-user latent variable $\pi_u$ (or functions of it, like
$\hat{a}_{ui}$) captures multi-treatment confounders, i.e., variables
that correlate with multiple exposures and the ratings vector (Lemma 3
of \citep{wang2018blessings}).  We note that the true confounding
mechanism does not need to coincide with \gls{PF} and nor does the
real confounder need to coincide with $\pi_u$. Rather, \gls{PF} produces
a substitute confounder that is sufficient to debias confounding. (We
formalize this justification in \Cref{sec:proof}.)

\parhead{Beyond probabilistic matrix factorization.} The deconfounder
involves two models, one for exposure and one for outcome. We have
introduced \gls{PF} as the exposure model and probabilistic matrix
factorization \citep{mnih2008probabilistic} as the outcome model.
Focusing on \gls{PF} as the exposure model, we can also use other
outcome models, e.g. Poisson matrix factorization
\citep{gopalan2015scalable} and weighted matrix factorization
\citep{hu2008collaborative}. We discuss these extensions in
\Cref{sec:generalMF}.

\begin{algorithm}[t]
  \DontPrintSemicolon
  \;
  \KwIn{a dataset of exposures and ratings $\{(a_{ui}, y_{ui}(a_{ui}))\}_{u,i}
    ,\,\, i=1, \ldots, I, u=1,\ldots,U$}

  \BlankLine

  \KwOut{the potential outcome given treatment $\hat{y}_{ui}(1)$}

  \BlankLine

  1. Fit \gls{PF} to the exposures $\{a_{ui}\}_{u,i}$ from \Cref{eq:pf}\;

  \BlankLine

  2. Compute substitute confounders $\{\hat{a}_{ui}\}_{u,i}$ from \Cref{eq:reconstr} \;

  \BlankLine

  3. Fit the outcome model $\{(a_{ui},
  y_{ui}(a_{ui}))\}_{u,i}$ from \Cref{eq:dcfrec} \;

  \BlankLine

  4. Estimate potential ratings $\hat{y}_{ui}(1)$ with the fitted outcome
  model (\Cref{eq:dcfrec})\;
  \caption{The Deconfounded Recommender}
  \label{alg:deconfounder}
\end{algorithm}


\section{Empirical Studies}

\label{sec:empirical}

We study the deconfounded recommender on simulated and real datasets.
We examine its recommendation performance and compare to existing
recommendation algorithms. We find that the deconfounded recommender
is more robust to unobserved confounding than existing approaches; it
predicts the ratings better and consistently improves
recommendation. (The supplement contains software that reproduces
these studies.)

\subsection{Evaluation of causal recommendation models}
\label{sec:howtoeval}

We first describe how we evaluate the recommender.  Recommender
systems are trained on a set of user ratings and tested on heldout
ratings (or exposures).  The goal is to evaluate the recommender with
the test sets.  Traditionally, we evaluate the accuracy (e.g. mean
squared error) of the predicted ratings.  Or we compute ranking
metrics: were the items with high ratings also ranked high in our
predictions?

However, causal recommendation models pose unique challenges for
evaluation. In causal inference, we need to evaluate how a model
performs across all potential outcomes,
\begin{align}
  \label{eq:causal-err}
  \textstyle \text{err}_{\textrm{cau}}
  =\frac{1}{U}\sum_{u=1}^U\ell(\{\hat{y}_{ui}\}_{i\in \{1,...,I\}},
  \{y_{ui}(1)\}_{i\in\{1,...,I\}}),
\end{align}
where $\ell$ is a loss function, such as \gls{MSE} or \gls{NDCG}.  The
challenge is that we don't observe all potential outcomes $y_{ui}(1)$.

Which test sets can we use for evaluating causal recommendation
models?  One option is to generate a test set by randomly splitting
the data; we call this a ``regular test set.'' However, evaluation on
the regular test set gives a biased estimate of
$\text{err}_{\textrm{cau}}$; it emphasizes popular items and active
users.

An (expensive) solution is to measure a randomized test set. Randomly
select a subset $\mathcal{I}_u$ from all items and ask the users to
interact and rate all of them. Then compute the average loss across
users,
\begin{align}
  \label{eq:randomized-err}
  \textstyle \text{err}_{\textrm{rand}}
  =\frac{1}{U}\sum_{u=1}^U\ell(\{\hat{y}_{ui}\}_{i\in \mathcal{I}_u},
  \{y_{ui}(1))\}_{i\in\mathcal{I}_u}).
\end{align}
\Cref{eq:randomized-err} is an unbiased estimate of the average across
all items in \Cref{eq:causal-err}; it tests the recommender's ability
to answer the causal question.  Two available datasets that include
such random test sets are the Yahoo! R3 dataset
\citep{marlin2009collaborative} and the coat shopping dataset
\citep{schnabel2016recommendations}. We also create random test sets
in simulation studies.

However, a randomized test set is often difficult to obtain. In this
case, our solution is to evaluate the average per-item predictive
accuracy on a ``regular test set.'' For each item, we compute the MSE
of all the ratings on this movie; we then average the MSEs of all
items. While popular items receive more ratings, this average per-item
predictive accuracy treats all items equally, popular or unpopular. We
use this metric to evaluate the deconfounded recommender on Movielens
100k and Movielens 1M datasets.~\looseness=-1


\subsection{Simulation studies}

We study the deconfounded recommender on simulated datasets. We
simulate movie ratings for $U=5,000$ users and $I=5,000$ items, where
effect of preferences on rating is confounded.

\parhead{Simulation setup.} We simulate a $K$-vector confounder for
each user $c_{u} \sim \mathrm{Gamma}_K(0.3, 0.3)$ and a
$K$-vector of attributes for each item
$\beta_i \sim \mathrm{Gamma}_K(0.3, 0.3)$. We then
simulate the user preference $K$-vectors $\theta_u$ conditional on the
confounders,
\begin{align*}
  \theta_u &\sim \gamma_\theta \cdot c_u +
  (1-\gamma_\theta) \cdot \mathrm{Gamma}_K(0.3, 0.3).
\end{align*}
The constant $\gamma_\theta\in [0,1]$ controls the exposure-confounder
correlation; higher values imply stronger confounding.

We next simulate the binary exposures $a_{ui}\in \{0,1\}$, the ratings
for all users watching all movies $y_{ui}(1)\in \{1,2,3,4,5\}$, and
calculate the observed ratings $y_{ui}$. The exposures and ratings are
both simulated from truncated Poisson distributions, and the observed
ratings mask the ratings by the exposure,\looseness=-1
\begin{align*}
  a_{ui}&\sim\min(\mathrm{Poisson}(c_u^\top \beta_i), 1),\\
  y_{ui}(1)&\sim \min(1+\mathrm{Poisson}((\theta_u +\gamma_y\cdot c_u)^\top \beta_i), 5),\\
  y_{ui} &= a_{ui} \cdot y_{ui}(1).
\end{align*}
The constant $\gamma_y\geq 0$ controls how much the confounder $c_u$
affects the outcome; higher values imply stronger confounding.

\parhead{Competing methods.} We compare the deconfounded recommender
to baseline methods. One set of baselines are the classical
counterparts of the deconfounded recommender. We explore probabilistic
matrix factorization \citep{mnih2008probabilistic}, Poisson matrix
factorization \citep{gopalan2015scalable}, and weighted matrix
factorization \citep{hu2008collaborative}; see
\Cref{sec:vanilladeconfounder} for details of these baseline models.
We additionally compare to \glsreset{IPW}\gls{IPW} matrix
factorization \citep{schnabel2016recommendations}, which also handles
selection bias in observational recommendation data.

\begin{figure}[t]
\hspace{-2pt}
\begin{minipage}[t]{\textwidth}
\begin{subfigure}[b]{0.33\textwidth}
\centering
\includegraphics[width=\textwidth]{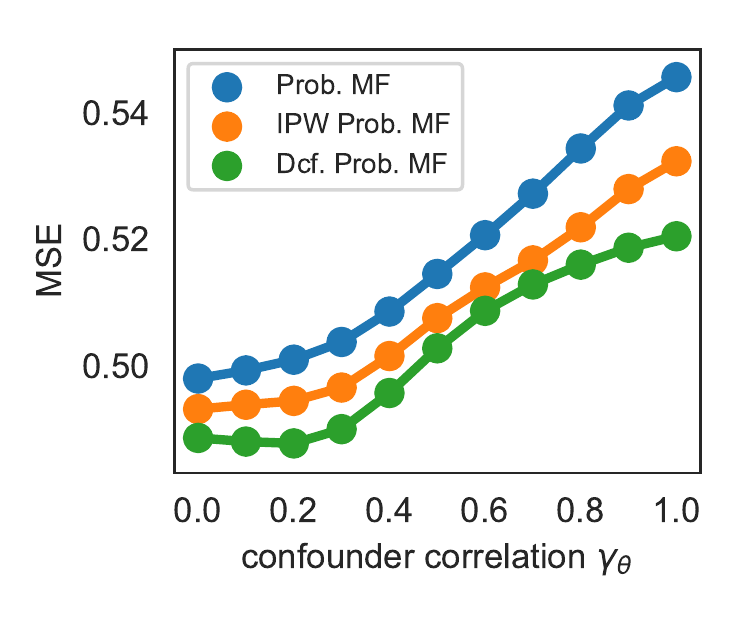}
\caption{Probabilistic MF
\label{fig:MSEconfstrength}}
\end{subfigure}
\begin{subfigure}[b]{0.33\textwidth}
\centering
\includegraphics[width=\textwidth]{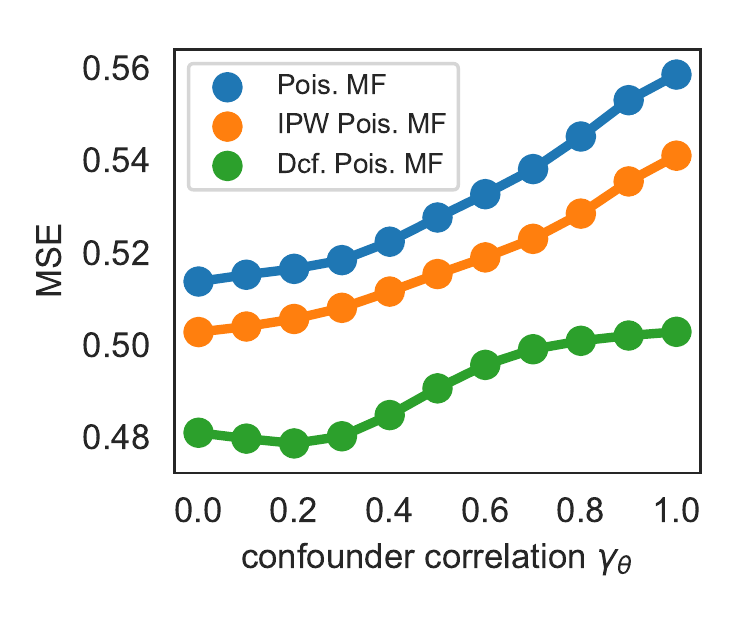}
\caption{Poisson MF\label{fig:MSEconfcorr}}
\end{subfigure}%
\begin{subfigure}[b]{0.33\textwidth}
\centering
\includegraphics[width=\textwidth]{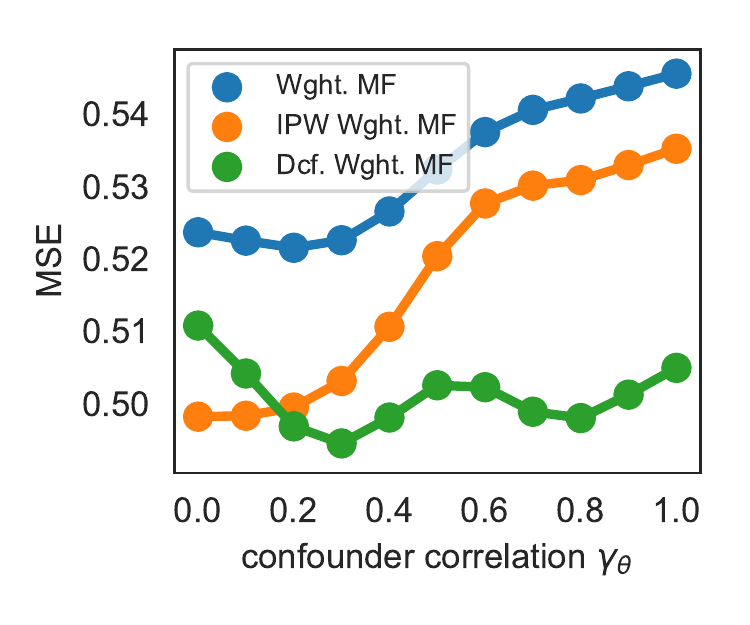}
\caption{Weighted MF\label{fig:NDCGconfstrength}}
\end{subfigure}%
\end{minipage}
\caption{\label{fig:varygamma1} Varying confounder correlation
$\gamma_\theta$ from 0.0 to 1.0 ($\gamma_y=3.0$). The rating
predictions of the deconfounded recommender (green) is more robust to
unobserved confounding than its classical counterpart (blue) and the
existing causal approach, IPW MF
\citep{schnabel2016recommendations} (orange). (Lower is better.)}
\end{figure}

\parhead{Results.}
\Cref{fig:varygamma1} shows how an unobserved confounder high
correlated with exposures can affect rating predictions. (Its effects
on ranking quality is in \Cref{sec:simfigs}.) Although the
performances of all algorithms degrades as the unobserved confounding
increases, the deconfounded recommender is more robust. It leads to
smaller MSEs in rating prediction and higher NDCGs in recommendation
than its classical counterparts (Probabilistic/Poisson/Weighted MF)
and the existing causal approach (IPW Probabilistic/Poisson/Weighted
MF) \citep{schnabel2016recommendations}.

\subsection{Case studies I: The deconfounded recommender on random
test sets}

We next study the deconfounded recommender on two real datasets:
Yahoo! R3 \citep{marlin2009collaborative} and coat shopping
\citep{schnabel2016recommendations}. Both datasets are comprised of an
observational training set and a random test set. The training set
comes from users rating user-selected items; the random test set comes
from the recommender system asking its users to rate randomly selected
items. The latter enables us to evaluate how different recommendation
models predict \textit{potential outcomes}: what would the rating be
if we \textit{make} a user watch and rate a movie?

\parhead{Datasets.}
Yahoo! R3 \citep{marlin2009collaborative} contains user-song ratings.
The training set contains over 300K user-selected
ratings from 15400 users on 1000 items. Its random test set contains 5400
users who were asked to rate 10 randomly chosen songs.
The coat shopping dataset \citep{schnabel2016recommendations} contains
user-coat ratings. The training set contains 290 users.
Each user supplies 24 user-selected ratings among 300 items. Its
random test contains ratings for 16 randomly selected coat per user.

\parhead{Evaluation metrics.} We use the recommenders for two types of
prediction: weak generalization and strong generalization
\citep{marlin2004modeling}. Weak generalization predicts preferences
of existing users in the training set on their unseen movies. Strong
generalization predicts preferences of new users---users not in the
training set---on their unseen movies. Based on the predictions, we
rank the items with nonzero ratings. To evaluate recommendation
performance, we report three standard measures: \gls{NDCG}, recall,
and MSE. See \Cref{sec:metrics} for formal definitions.

\parhead{Experimental setup.} For each dataset, we randomly split
80/20 the training set into training/validation sets. We leave the
random test set intact. Across all experiments, we use the validation
set to select the best hyperparameters for the recommendation models.
We choose the hyperparameters that yield the best validation log
NDCG. The latent dimension is chosen from
$\{1,2,5,10,20,50,100\}$.

The deconfounded recommender has two components: the treatment
assignment model and the outcome model. We always use \gls{PF} as the
treatment assignment model. The hyperparameters of both models are
chosen together based on the same validation set as above. The best
hyperparameters are those that yield the best validation NDCG.

\parhead{Results.} \Cref{tab:wg,tab:sg} show the recommendation
performance of the deconfounded recommender and its
competitors. Across the three metrics and the two datasets, the
deconfounded recommender outperforms its classical counterpart for
both weak and strong generalization: it produces better item rankings
and improves retrieval quality; its predicted ratings are also more
accurate. The deconfounded recommender also outperforms the
\gls{IPW} matrix factorization
\citep{schnabel2016recommendations}, which is the main existing
approach that targets selection bias in recommendation systems. These
results show that the deconfounded recommender produces more accurate
predictions of user preferences by accommodating unobserved
confounders in item exposures.


\begin{table*}
\footnotesize
  \begin{center}
    \begin{tabular}{lcccp{0.01mm}ccc} 
     \toprule
       &  \multicolumn{3}{c}{\textbf{Yahoo! R3}} & &\multicolumn{3}{c}{\textbf{Coat}} \\
       & NDCG & Recall@5 & MSE &  &NDCG & Recall@5 & MSE\\
          \midrule
          Probabilistic MF [\citenum{mnih2008probabilistic}] 
          &0.772 & 0.540 &1.963&&0.728 & 0.552& 1.422\\
          IPW Probabilistic MF [\citenum{schnabel2016recommendations}] 
          &0.791 & 0.603 &1.893&&0.732 & 0.547 & 1.358\\
          Deconfounded Probabilistic MF 
          &\bfseries{0.819} & \bfseries{0.640} &\bfseries{1.768}&&\bfseries{0.743} & \bfseries{0.569} &\bfseries{1.341}\\
          \cdashlinelr{1-8}
          Poisson MF [\citenum{gopalan2015scalable}]
          &0.789 & 0.564& 1.539&&0.713 &0.451 &1.657\\
          IPW Poisson MF [\citenum{schnabel2016recommendations}] 
          &0.792 & 0.564 & 1.513 && 0.693 & 0.458 & \bfseries{1.631} \\
          Deconfounded Poisson MF 
          &\bfseries{0.802}  &\bfseries{0.610} &\bfseries{1.447}&&\bfseries{0.743} &\bfseries{0.521 }&1.657\\
          \cdashlinelr{1-8}
          Weighted MF [\citenum{hu2008collaborative}] 
          &0.820 & 0.639 &2.047&&0.738 & 0.560& 1.658\\
          IPW Weighted MF [\citenum{schnabel2016recommendations}] 
          & 0.804& 0.592 & 1.845 &&0.681  & 0.554 & 1.612\\
          Deconfounded Weighted MF 
          &\bfseries{0.823} &\bfseries{0.645} &\bfseries{1.658}&&\bfseries{0.744} & \bfseries{0.581}& \bfseries{1.569}\\ 
      \bottomrule
    \end{tabular}
    \caption{Recommendation on random test sets for existing users
    (weak generalization). The deconfounded recommender improves
    recommendation over classical approaches and the existing causal
    approach \citep{schnabel2016recommendations}. (Higher is
    better for NDCG and Recall@5; lower is better for MSE.)\label{tab:wg}}
  \end{center}
\end{table*}


\begin{table*}
\footnotesize
  \begin{center}
    \begin{tabular}{lcccp{0.01mm}ccc} 
     \toprule
       &  \multicolumn{3}{c}{\textbf{Yahoo! R3}} && \multicolumn{3}{c}{\textbf{Coat}} \\
       & NDCG & Recall@5 & MSE &  &NDCG & Recall@5 & MSE\\
          \midrule         
          Probabilistic MF [\citenum{mnih2008probabilistic}] 
          &0.802 &0.792 &2.307&&0.833 & 0.737& \bfseries{1.490}\\
          IPW Probabilistic MF [\citenum{schnabel2016recommendations}] 
          &0.818 &0.827 &2.625&&0.811  & 0.724 & 1.587\\
          Deconfounded Probabilistic MF 
          &\bfseries{0.824} & \bfseries{0.829} &\bfseries{2.244}&&\bfseries{0.847}& \bfseries{0.808} & 1.540\\
          \cdashlinelr{1-8}
          Poisson MF [\citenum{gopalan2015scalable}] 
          &0.765& 0.752& 1.913&&0.764 & 0.671& 1.889\\
          IPW Poisson MF [\citenum{schnabel2016recommendations}] 
          &0.769 & 0.761 & \bfseries{1.876} && 0.769 & \bfseries{0.678} & 1.817 \\
          Deconfounded Poisson MF 
          &\bfseries{0.774}  &\bfseries{0.769} &1.881&&\bfseries{0.772}& \bfseries{0.678} & \bfseries{1.811}\\
          \cdashlinelr{1-8}
          Weighted MF [\citenum{hu2008collaborative}] 
          &0.809 & 0.793& 1.904&&0.850 &0.814&2.859\\
          IPW Weighted MF [\citenum{schnabel2016recommendations}] 
          &0.786 & 0.788 & 1.883 && 0.837 & 0.800 &2.477\\
          Deconfounded Weighted MF 
          &\bfseries{0.820} &\bfseries{0.818}& \bfseries{1.644}&&\bfseries{0.854} &\bfseries{0.829}& \bfseries{2.421}\\           
      \bottomrule
    \end{tabular}
    \caption{Recommendation on random test sets for new users (strong
    generalization). The deconfounded recommender improves
    recommendation over classical approaches and the existing causal
    approach \citep{schnabel2016recommendations}. (Higher is
    better for NDCG and Recall@5; lower is better for MSE.)\label{tab:sg}}
  \end{center}
\end{table*}

\subsection{Case studies II: The deconfounded recommender on
regular test sets}

\label{subsec:causalvsnoncausal}

We study the deconfounded recommender on two MovieLens datasets:
ML100k and ML1M.\footnote{http://grouplens.org/datasets/movielens/}
These datasets only involve observational data; they do not contain
randomized test sets. We focus on how well we predicts on \emph{all}
items, popular or unpopular.

\parhead{Datasets.} The Movielens 100k dataset contains 100,000
ratings from 1,000 users on 1,700 movies. The Movielens 1M dataset
contains 1 million ratings from 6,000 users on 4,000 movies.

\parhead{Experimental setup and performance measures.} We employ the
same experimental protocols for hyperparameter selection as before.
For each dataset, we randomly split the training set into
training/validation/test sets with 60/20/20 proportions. We measure
the recommendation performance by the average per-item predictive
accuracy, which equally treats popular and unpopular items; see
\Cref{sec:howtoeval} for details.

\parhead{Results.} \Cref{tab:regular} presents the recommendation
performance of the deconfounded recommender and their classical
counterpart on average per-item MSEs and MAEs. Across the two metrics
and two datasets, the deconfounded recommender leads to lower MSEs and
MAEs on predictions over all items than classical approaches and the
existing causal approach, IPW MF~\citep{schnabel2016recommendations}.
Instead of focusing on popular items, the deconfounded recommender
targets accurate predictions on \emph{all} items. Hence it improves
the prediction quality across all items.


\begin{table*}
\footnotesize
  \begin{center}
    \begin{tabular}{lccp{0.01mm}cc} 
     \toprule
       &  \multicolumn{2}{c}{\textbf{Movielens 100K}} && \multicolumn{2}{c}{\textbf{Movielens 1M}} \\
       & MSE & MAE   && MSE  & MAE \\
          \midrule
          Probabilistic MF [\citenum{mnih2008probabilistic}] 
          & 2.926& 1.425&& 2.774 & 1.321\\
          IPW Probabilistic MF [\citenum{schnabel2016recommendations}] 
          & 2.609& 1.275&& 2.714 & 1.303\\
          Deconfounded Probabilistic MF
          & \bfseries{2.554}& \bfseries{1.260}&& \bfseries{2.699} & \bfseries{1.299}\\
          \cdashlinelr{1-6}
          Poisson MF [\citenum{gopalan2015scalable}] 
          & 3.374& 1.475&& 2.357 & 1.305\\
          IPW Poisson MF [\citenum{schnabel2016recommendations}] 
          & 3.439& 1.480&& \bfseries{2.196} & \bfseries{1.220}\\
          Deconfounded Poisson MF 
          & \bfseries{3.268}& \bfseries{1.454}&& 2.325 & 1.295\\
          \cdashlinelr{1-6}
          Weighted MF [\citenum{hu2008collaborative}] 
          & 2.359& 1.219&& 3.558 & 1.516\\
          IPW Weighted MF [\citenum{schnabel2016recommendations}] 
          & 2.344& 1.198&& 2.872 & 1.363\\
          Deconfounded Weighted MF  
          & \bfseries{2.101}& \bfseries{1.138}&& \bfseries{2.864} & \bfseries{1.360}\\
      \bottomrule
    \end{tabular}
    \caption{Average per-item predictive accuracy on heldout ratings.
The deconfounded recommender leads to lower MSEs and MAEs on
predictions over all items; it outperforms classical approaches and
the existing causal approach \citep{schnabel2016recommendations}.
(Lower is better.)
    \label{tab:regular}}
  \end{center}
\end{table*}


\section{Discussion}
\label{sec:discussion}

We develop the \textit{deconfounded recommender}, a strategy to use
classical recommendation models for causal predictions: how would a
user rate a recommended movie? The deconfounded recommender uses
Poisson factorization to infer confounders in treatment assignments;
it then augments common recommendation models to correct for
confounding bias. The deconfounded recommender improves recommendation
performance and rating predictions; it is also more robust to
unobserved confounding in user exposures.

\clearpage
{\small\putbib[BIB1]}
\end{bibunit}

\clearpage
\begin{bibunit}[alp]

\appendix

\onecolumn

{\Large\textbf{Supplementary Material}}

\section{Theoretical justification of the deconfounded recommender}
\label{sec:proof}

The following theorem formalizes the theoretical justification of the
deconfounded recommender.
\begin{thm}
  If no confounders affect the exposure to only one of the items, and
  the number of items go to infinity, then the deconfounded
  recommender forms unbiased causal inference
  \begin{align}
  \label{eq:identification}
  E[Y_{ui}(a)] = E[E[Y_{ui}(A_{ui})\g A_{ui}=a, \pi_u^\top \lambda_i]] \quad
    \text{ for all }u,i,
  \end{align}
  when $A_{ui}\sim \textrm{Poisson}(\pi_u^\top
  \lambda_i)$ for some independent random vectors $\pi_u$'s and $\lambda_i$'s,
\end{thm}
This theorem follows from Proposition 5 and Theorem 7 of
\citep{wang2018blessings} in addition to the fact that $\pi_u
\independent \mathbb{1}\{A_{ui}=a\} \g p(A_{ui}=a\g \pi_u^\top \lambda_i)$
\citep{hirano2004propensity}. Specifically, Proposition 5 of
\citep{wang2018blessings} asserts the ignorability of the substitute
confounder; Theorem 7 of \citep{wang2018blessings} establishes causal
identification of the deconfounder on the intervention of subsets of
the causes. The deconfounded recommender fits into this setting
because $Y_{ui}(a)$ is a counterfactual quantity involving only one
item (cause), i.e. the exposure to item $i$.

Proposition 5 and Theorem 7 of \citet{wang2018blessings} implies
\[E[Y_{ui}(a)] = E[E[Y_{ui}(A_{ui})\g A_{ui}=a, \pi_u]] \quad \text{
for all  }u,i.\] Therefore, we have
\begin{align*}
E[Y_{ui}(a)] &= E[E[Y_{ui}(A_{ui})\g A_{ui}=a, \pi_u]]\\
&=E[E[Y_{ui}(A_{ui})\g \mathbb{1}\{A_{ui}=a\}, \pi_u]]\\
&=E[E[E[Y_{ui}(A_{ui})\g \mathbb{1}\{A_{ui}=a\}, \pi_u, p(A_{ui}=a\g \pi_u^\top \lambda_i)]]]\\
&= E[E[E[Y_{ui}(A_{ui})\g \mathbb{1}\{A_{ui}=a\}, p(A_{ui}=a\g \pi_u^\top \lambda_i)]]]\\
&= E[E[E[Y_{ui}(A_{ui})\g \mathbb{1}\{A_{ui}=a\}, \pi_u^\top \lambda_i]]]\\
&= E[E[E[Y_{ui}(A_{ui})\g A_{ui}=a, \pi_u^\top \lambda_i]]]\\
\end{align*}

The first three equalities are due to basic probability facts. The
fourth equality is due to $\pi_u \independent
\mathbb{1}\{A_{ui}=a\} \g p(A_{ui}=a\g \pi_u^\top \lambda_i)$ \citep{hirano2004propensity}. 
The fifth equality is due to $A_{ui}\sim\textrm{Poisson}(\pi_u^\top
\lambda_i)$. The last equality is again due to basic probability facts.

\section{Beyond probabilistic matrix factorization}
\label{sec:generalMF}

Focusing on \gls{PF} as the exposure model, we extend the deconfounded
recommender to general outcome models.

We start with a general form of matrix factorization,
\begin{align}
  \label{eq:general-model}
  y_{ui}(a)\sim p(\cdot\g m(\theta_u^\top \beta_i, a), v(\theta_u^\top
  \beta_i, a)),
\end{align}
where $m(\theta_u^\top \beta_i, a)$ characterizes the mean and
$v(\theta_u^\top \beta_i, a)$ the variance of the ratings
$y_{ui}(a)$. This form encompasses many factorization models.
Probabilistic matrix factorization \citep{mnih2008probabilistic} is
\[m(\theta_u^\top \beta_i, a) = a\cdot \theta_u^\top \beta_i , \quad
  v(\theta_u^\top \beta_i, a) = \sigma^2,\] and $p(\cdot)$ is the
Gaussian distribution. Weighted matrix factorization
\citep{hu2008collaborative} also involves a Gaussian $p$, but its
variance changes based on whether a user has seen the movie:
\[m(\theta_u^\top \beta_i, a) =
\theta_u^\top \beta_i , \quad v(\theta_u^\top \beta_i, a) =
\sigma^2_{a},\]
where $\sigma_0^2=\alpha\sigma_1^2$. This model leads us to downweight
the zeros; we are less confident about the zero ratings. Poisson
matrix factorization as an outcome model \citep{gopalan2015scalable}
only takes in a mean parameter
\[m(\theta_u^\top \beta_i, a) =
\theta_u^\top \beta_i\]
and set $p(\cdot)$ to be the Poisson distribution.

With the general matrix factorization of \Cref{eq:general-model}, the
deconfounded recommender fits an augmented outcome model $M_Y$.  This
outcome model $M_Y$ includes the substitute confounder as a covariate,
\begin{align}
\label{eq:dcfrec}
y_{ui}(a) \sim p(\cdot\g m(\theta_u^\top \beta_i, a)+\gamma_u\hat{a}_{ui}+\beta_0, v(\theta_u^\top \beta_i, a)).
\end{align}
Notice the parameter $\gamma_u$ is a user-specific coefficient; for
each user, it characterizes how much the substitute confounder
$\hat{a}$ contributes to the ratings. Note the deconfounded
recommender also includes an intercept $\beta_0$. These deconfounded
outcome model can be fit by \textit{maximum a posteriori} estimation.

It solves
\begin{align*}
&\textstyle \hat{\theta}_u, \hat{\beta}_i, \hat{\gamma}_u, \hat{\beta}_0 \\
= &\argmax
\sum_{u=1}^U\sum_{i=1}^I \log p(y_{ui}; m(\theta_u^\top \beta_i,
a_{ui})+\gamma_u\hat{a}_{ui}+\beta_0, \\
&v(\theta_u^\top \beta_i, a_{ui})) \textstyle +
\sum_{u=1}^U\log p(\theta_u)+\sum_{i=1}^I \log p(\beta_i)\\
&+\sum_{u=1}^U\log
p(\gamma_u)+\log p(\beta_0),
\end{align*}
where $p(\theta_u)$, $p(\beta_i)$, $p(\gamma_u)$, and $p(\beta_0)$ are
priors of the latent variables.

To form recommendations, the deconfounded recommender predicts all of
the potential ratings, $y_{ui}(1)$. For an existing user $u$, it
computes the potential ratings from the fitted outcome model,
\begin{align}
\label{eq:pred1}
\hat{y}_{ui}(1) =m(\hat{\theta}_u^\top \hat{\beta}_i, 1)+\hat{\gamma}_u\cdot\hat{a}_{ui}+\hat{\beta}_0.
\end{align}
For a new user $u'$ with only a few ratings, it fixes the item vectors
$\lambda_i$ and $\beta_i$, and compute user vectors for the new user: it
fits $\pi_{u'}$, $\theta_{u'}$, and $\gamma_{u'}$ from the exposure and
the ratings of this new user $u'$. It finally computes the prediction,
\begin{align}
\label{eq:pred2}
\hat{y}_{u'i}(1) =m(\hat{\theta}_{u'}^\top \hat{\beta}_i, 1)+\hat{\gamma}_{u'}\cdot\hat{\pi}_{u'}^\top \lambda_i +\hat{\beta}_0.
\end{align}
The deconfounded recommender ranks all the items for each user based
on $\hat{y}_{ui}(1), i=1,\ldots,I$, and recommends highly ranked
items.

\section{Detailed simulation results}

\label{sec:simfigs}

\begin{figure}[t]
\hspace{-2pt}
\begin{minipage}[t]{\textwidth}
\begin{subfigure}[b]{0.33\textwidth}
\centering
\includegraphics[width=\textwidth]{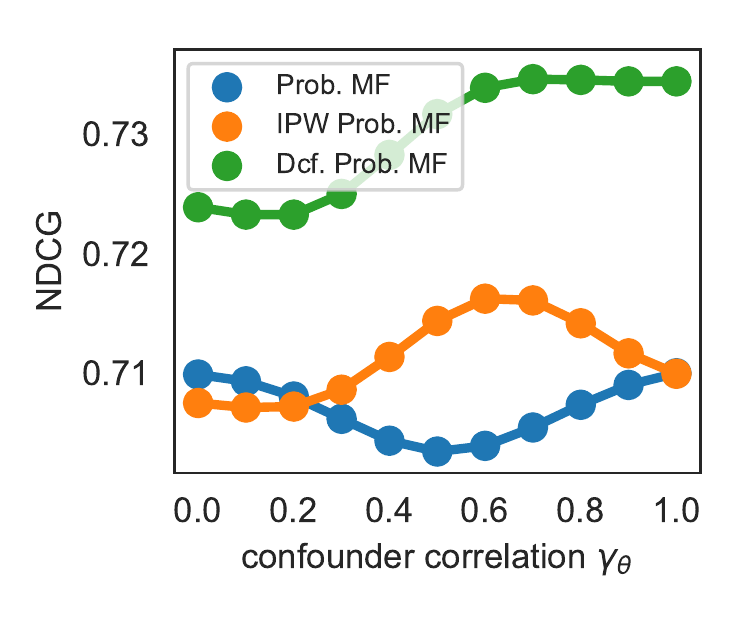}
\caption{Probabilistic MF
\label{fig:MSEconfstrength}}
\end{subfigure}
\begin{subfigure}[b]{0.33\textwidth}
\centering
\includegraphics[width=\textwidth]{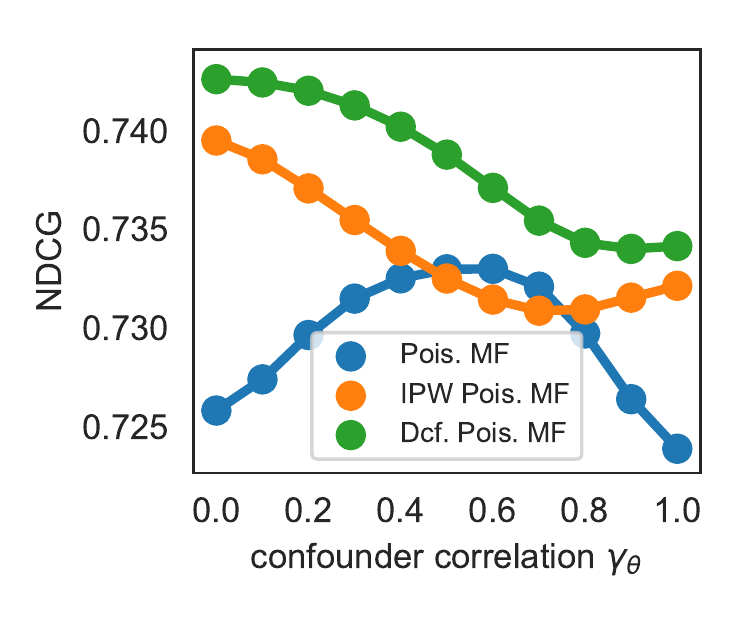}
\caption{Poisson MF\label{fig:MSEconfcorr}}
\end{subfigure}%
\begin{subfigure}[b]{0.33\textwidth}
\centering
\includegraphics[width=\textwidth]{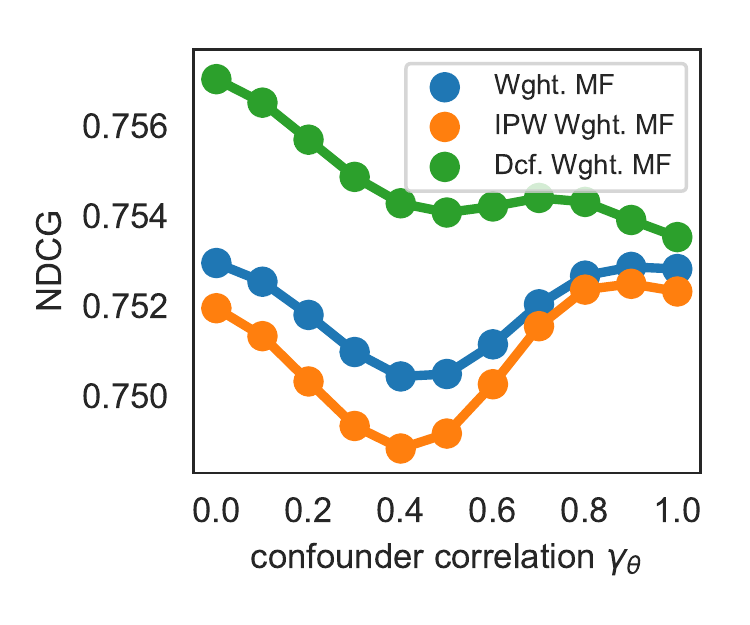}
\caption{Weighted MF\label{fig:NDCGconfstrength}}
\end{subfigure}%
\end{minipage}
\caption{\label{fig:varygamma2} Varying confounding correlation
$\gamma_\theta$ from 0.0 to 1.0 ($\gamma_y=3.0$). The ranking quality
of the deconfounded recommender (green) is more robust to unobserved
confounding than its classical counterpart (blue) and the existing
causal approach (orange) \citep{schnabel2016recommendations}. Higher
is better. (Results averaged over 10 runs.)}
\end{figure}

\Cref{fig:varygamma2,fig:varygamma1} explore how an unobserved
confounder high correlated with exposures can affect rating
predictions and ranking quality of recommendation algorithms.

\begin{figure}[t]
\hspace{-2pt}
\begin{minipage}[t]{\textwidth}
\begin{subfigure}[b]{0.33\textwidth}
\centering
\includegraphics[width=\textwidth]{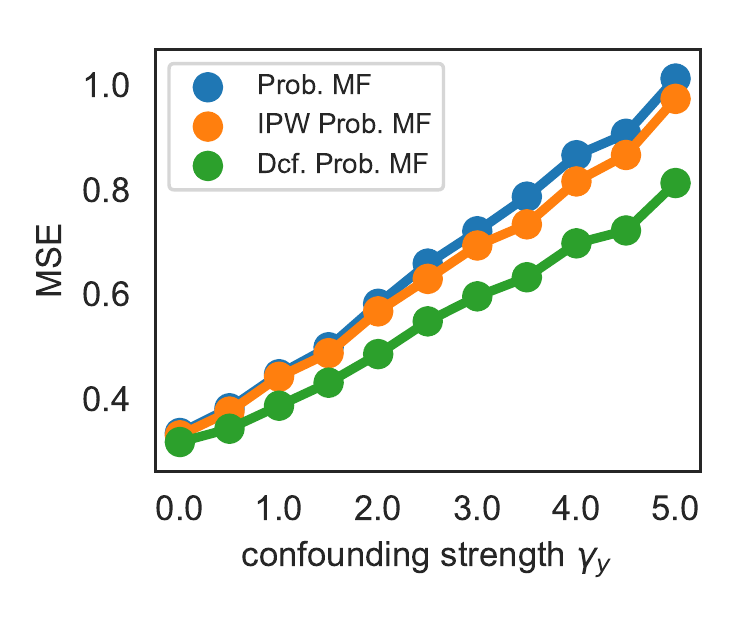}
\caption{Probabilistic MF
\label{fig:MSEconfstrength}}
\end{subfigure}
\begin{subfigure}[b]{0.33\textwidth}
\centering
\includegraphics[width=\textwidth]{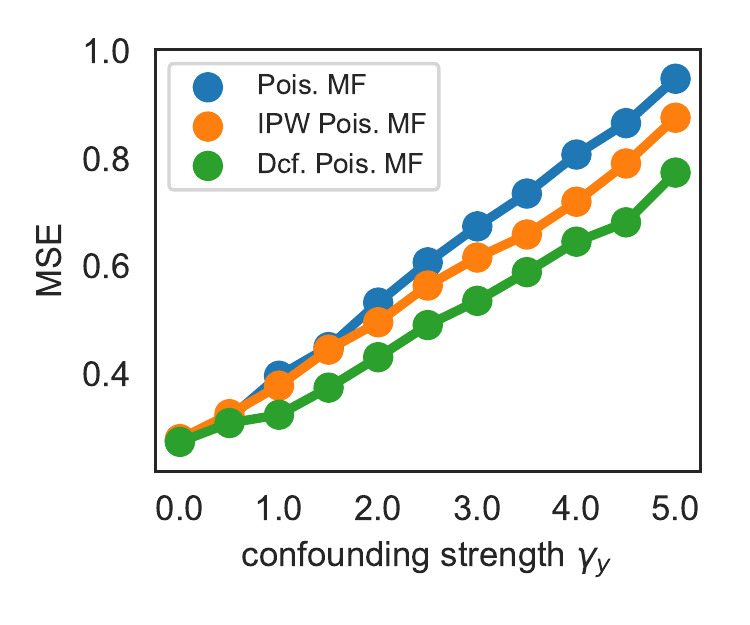}
\caption{Poisson MF\label{fig:MSEconfcorr}}
\end{subfigure}%
\begin{subfigure}[b]{0.33\textwidth}
\centering
\includegraphics[width=\textwidth]{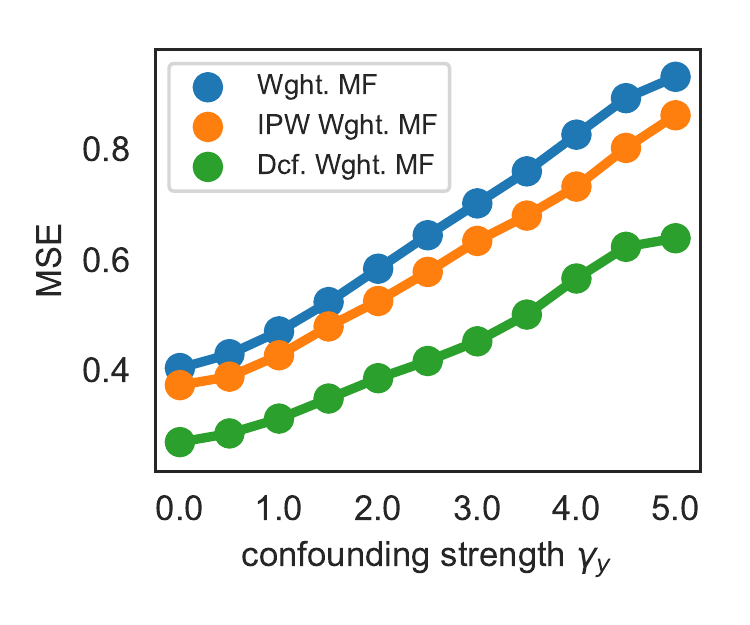}
\caption{Weighted MF\label{fig:NDCGconfstrength}}
\end{subfigure}%
\end{minipage}
\caption{\label{fig:varygamma3} Varying confounding strength
$\gamma_y$ from 0.0 to 5.0 ($\gamma_\theta=0.6$). The rating
predictions of the deconfounded recommender (green) is more robust to
unobserved confounding than its classical counterpart (blue) and the
existing causal approach (orange) \citep{schnabel2016recommendations}.
Lower is better. (Results averaged over 10 runs.)}
\end{figure}

\begin{figure}[t]
\hspace{-2pt}
\begin{minipage}[t]{\textwidth}
\begin{subfigure}[b]{0.33\textwidth}
\centering
\includegraphics[width=\textwidth]{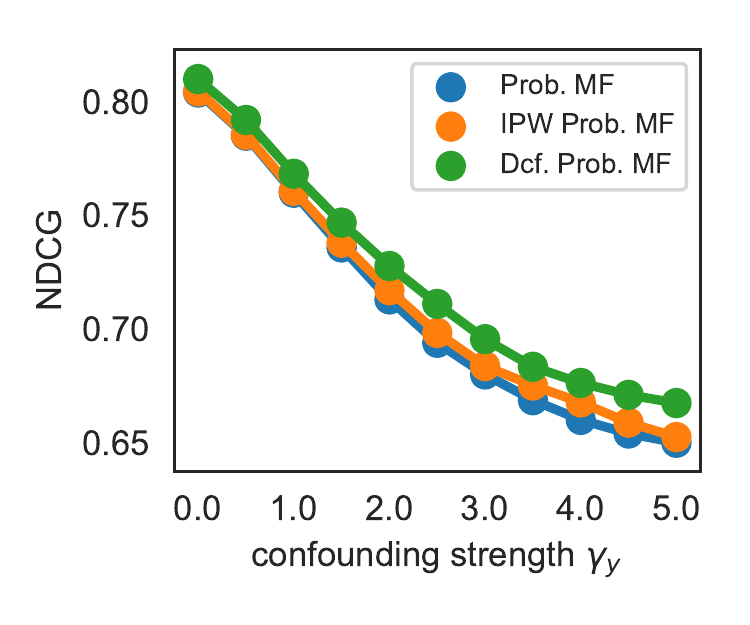}
\caption{Probabilistic MF
\label{fig:MSEconfstrength}}
\end{subfigure}
\begin{subfigure}[b]{0.33\textwidth}
\centering
\includegraphics[width=\textwidth]{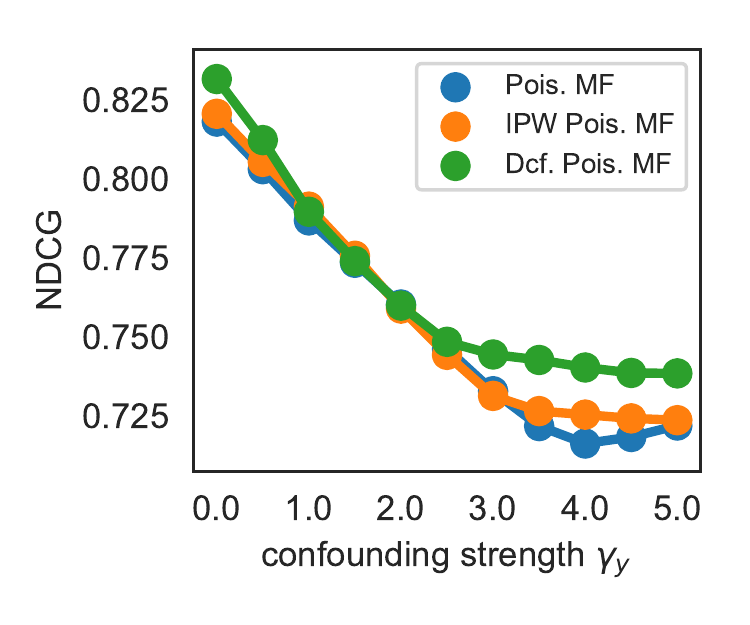}
\caption{Poisson MF\label{fig:MSEconfcorr}}
\end{subfigure}%
\begin{subfigure}[b]{0.33\textwidth}
\centering
\includegraphics[width=\textwidth]{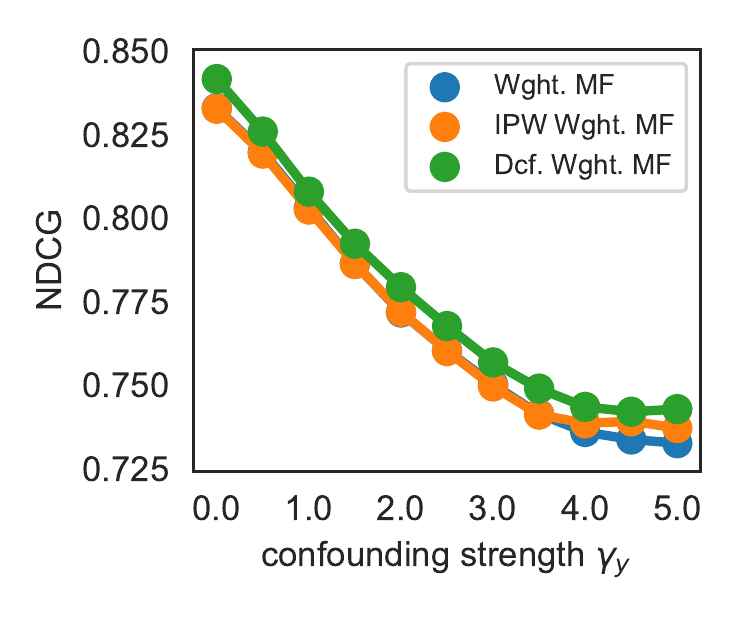}
\caption{Weighted MF\label{fig:NDCGconfstrength}}
\end{subfigure}%
\end{minipage}
\caption{\label{fig:varygamma4} Varying confounding strength
$\gamma_y$ from 0.0 to 5.0 ($\gamma_\theta=0.6$).  The ranking quality
of the deconfounded recommender (green) is more robust to unobserved
confounding than its classical counterpart (blue) and the existing
causal approach (orange) \citep{schnabel2016recommendations}. Higher
is better. (Results averaged over 10 runs.)}
\end{figure}

\Cref{fig:varygamma3,fig:varygamma4} explore how an unobserved
confounder high correlated with the rating outcome can affect rating
predictions and ranking quality of recommendation algorithms.

In both simulations, we find that the deconfounded recommender is more
robust to unobserved confounding; it leads to smaller MSEs in rating
prediction and higher NDCGs in recommendation than its classical
counterparts (Probabilistic/Poisson/Weighted MF) and the existing
causal approach (\gls{IPW} MF \citep{schnabel2016recommendations}).

\section{Performance measures}
\label{sec:metrics}
Denote $rank(u,i)$ as the rank of item $i$ in user $u$'s predicted
list; let $y_{u}^{\mathrm{test}}$ as the set of relevant items in the
test set of user $u$.\footnote{We consider items with a rating greater
than or equal to three as relevant.}
\begin{itemize}
  \item 
  \gls{NDCG} measures ranking quality:
  \[DCG = \frac{1}{U}\sum_{u=1}^U\sum_{i=1}^{L}\frac{2^{\mathrm{rel}_{ui}-1}}{\log_2(i+1)}, \quad NDCG = \frac{DCG}{IDCG},
\]
  where $L$ is the number of items in the test set of user $u$, the
  relevance $\mathrm{rel}_{ui}$ is set as the rating of user $u$ on
  item $i$, and $IDCG$ is a normalizer that ensures $NDCG$ sits
  between 0 and 1.
  \item Recall@k. Recall@k evaluates how many relevant items are
  selected:
  \[
  \textrm{Recall@k} = \frac{1}{U}\sum_{u=1}^U\sum_{i\in y_{u}^{\mathrm{test}}}\frac{\mathbb{1}\{rank(u,i)\leq k\}}{\min (k, |y_{u}^{\mathrm{test}}|)}.
  \]
 \item \textrm{MSE}. MSE evaluates how far the predicted ratings are
 away from the true rating:
  \[
  \textrm{MSE} = \frac{1}{|(u,i)\in y^{\mathrm{test}}|}\sum_{(u,i)\in
  y^{\mathrm{test}}} (\hat{y}_{ui} - y_{ui})^2.
  \]
\end{itemize}

\section{Experimental Details}

For weighted matrix factorization, we set weights of the observation
by $c_{ui} = 1 + \alpha y_{ui}$ where $\alpha = 40$ following
\citep{hu2008collaborative}.

For Gaussian latent variables in probabilistic matrix factorization,
we use priors $\cN(0,1)$ or $\cN(0,0.1^2)$.

For Gamma latent variables in \gls{PF}, we use prior
$\textrm{Gamma}(0.3, 0.3).$

For \glsreset{IPW}\gls{IPW} matrix factorization
\citep{schnabel2016recommendations}, we following Section 6.2 of
\citet{schnabel2016recommendations}. We set the propensity $p_{ui}$ to
be $p_{ui} = k$ for ratings $y_{ui}\geq 4$ and $p_{ui} =
k\alpha^{4-y_{ui}}$ for ratings $y_{ui}<4$, where $\alpha=0.25$ and
$k$ is set so that $\frac{1}{U\cdot I} \sum_{u=1}^U\sum_{i=1}^I p_{ui}
=
0.05$. We then perform \glsreset{IPW}\gls{IPW} matrix factorization with
the estimated propensities.
\clearpage
\putbib[BIB1]
\end{bibunit}

\end{document}